# Temporal Clustering Network for Self-Diagnosing Faults from Vibration Measurements

Geng Zhang
Michigan Engineering Services
gengz@miengsrv.com

Adam R. Singer
Michigan Engineering Services
arsinger@miengsrv.com

Nickolas Vlahopoulos
University of Michigan
nickvl@umich.edu

## ABSTRACT

There is a need to build intelligence in operating machinery and use data analysis on monitored signals in order to quantify the health of the operating system and self-diagnose any initiations of fault. Built-in control procedures can automatically take corrective actions in order to avoid catastrophic failure when a fault is diagnosed. This paper presents a Temporal Clustering Network (TCN) capability for processing acceleration measurement(s) made on the operating system (i.e. machinery foundation, machinery casing, etc.), or any other type of temporal signals, and determine based on the monitored signal when a fault is at its onset. The new capability uses: one-dimensional convolutional neural networks (1D-CNN) for processing the measurements; unsupervised learning (i.e. no labeled signals from the different operating conditions and no signals at pristine vs. damaged conditions are necessary for training the 1D-CNN); clustering (i.e. grouping signals in different clusters reflective of the operating conditions); and statistical analysis for identifying fault signals that are not members of any of the clusters associated with the pristine operating conditions. A case study demonstrating its operation is included in the paper. Finally topics for further research are identified.

## 1. Introduction

The development and wide application of Machine Learning (ML) algorithms based on Convolutional Neural Networks (CNN) has led to applying them for detection of faults from vibration measurements on operating machinery and structures [1-11]. These efforts recognize the value of building intelligence in operating machinery/structures and using data analysis on the collected signals in order to quantify the health of the operating system and self-diagnose any initiation of fault. As an example of specific engineering applications, autonomous maritime vehicles have a particular need for self-diagnosing faults for both Defense and Commercial autonomous vessels. Depending on the intended mission of the unmanned vehicles the speed requirements and thus the load intensity of the systems varies. For example, unmanned Naval vehicles [12] that serve as decoys for manned vessels must be able to travel at speeds consistent with the platforms whose signatures they simulate. Platforms tracking adversary ships or submarines must also travel at high enough speeds to keep pace with the vessels that they are tracking. The situation is different in sensing missions, such as minehunting, since speed is less of a consideration, but even then, failures can take place much more frequently than desired (cancellation of the Remote Multi-Mission Vehicle which exhibited ~19 hours between failures instead of the expected 75 hours) [13]. Therefore, it is necessary to be able to self-diagnose impeding failure from easy to collect measurements, in order to initiate automatically corrective actions that will avoid catastrophic failure (for example reduce the power of an operating propeller when fault in the bearing is identified, etc.). Such measures will allow the vehicle to continue operation, perhaps at a reduced speed, until a manual maintenance intervention can take place. As the commercial shipping Industry is also investing in unmanned ships for future operations [14, 15], there is great interest for self-diagnosing faults from easy to collect vibration measurements and preventing catastrophic failures.

The majority of reported work that uses CNN for fault identification from measured vibration time domain signals utilizes one-dimensional CNN (1D-CNN). A comprehensive review of 1D-CNN is presented in [1]; it offers technical background about their operation, their training, and their predictive capabilities, along with a summary of the technical areas where they have been used. In previous work supervised training has been employed for training the 1D-CNN to recognize and classify correctly time domain measurements which originate from both pristine and from various states of damage for all operating conditions of interest. The work presented here is based on the concept of the Deep Temporal Clustering presented in [16]. In our work, unlabeled time signals from all operating conditions at a pristine state are expected to be available and used for unsupervised training of a Temporal Clustering Network (TCN). This expectation is reasonable, because acceleration measurements can be collected during the trials of a new system or during its early operational stages, when the system is still at its pristine condition. These measurements will be used for training. Once the unsupervised training has been completed, the TCN can then process any candidate signal and identify if it corresponds to a fault state or if it is associated with one of the expected operating conditions.

The TCN training considers two targets (i.e. minimizing a reconstruction loss and minimizing a clustering loss), similar to the Deep Temporal Clustering presented in [16]. The latter comprises an unsupervised method for clustering time domain signals. Certain elements are modified in the process and new elements are introduced in this work based on specifically targeting the application of recognizing faults from vibration measurements. In the TCN the optimization for the reconstruction loss function and for the clustering are performed sequentially instead of jointly; the Bi-LSTM layers are not included in the encoder, that changes the nature of the features which are used for clustering; the k-mean algorithm is used for the initial clustering instead of a hierarchical algorithm; and a different loss function is defined for the final clustering. Finally, the concepts of threshold probability and failure probability are introduced for recognizing the fault signals. Brief background on 1D-CNN is presented first as a popular approach for supervised training in recognizing vibration signals. The structure of the TCN (it is a three step process) and the associated mathematical formulation is presented. The implementation is done using Python based libraries functions (PyTorch and Scikit-Learn). Acceleration measurements from an electric motor which are available from [17] are used in order to demonstrate how the method operates. Several steps which will be necessary in further developing/validating this method are identified.

**2. 1D-CNN**

In this Section brief technical background on 1D-CNN is presented since it comprises the primary approach utilized in the past for identifying faults by processing vibration measurements in a supervised manner. A useful technical discussion on 1D-CNN is presented in [18], and Figure 1 uses material from [18] in presenting the operational structure of 1D-CNN.

A time domain signal (in this case measured vibration) comprises the input. At each convolutional layer the time domain signal gets convolved with the number of filters assigned to the layer. The results from each convolution get down sampled through pooling (i.e. larger values are kept and the dimension of the signal gets reduced) and the outcome comprises the input to the next layer where a new set of convolutional and pooling operations are performed. The outcome from the final convolutional layer is processed using the filter weights of the classifier in order to generate the final classification decision (i.e. determine the label associated with the signal and identify what the signal is). In order to perform this process it is necessary to have weights for the filters of the various convolutional layers (feature extractor) and also the filter weights of the classifier. Both sets of these weights are determined during the training before the 1D-CNN can be used. In order to perform the training a large number of labeled training data must be available. This means that a large set of time domain signals along with the correct label (i.e. what the signal is) must be available. During the training a loss function is minimized by selecting appropriate values for filter weights in both the feature extractor and the classifier. The loss function is defined mathematically as the cumulative error between the label that is determined by the 1D-CNN and the actual known label of the input time domain signal, over all the labeled signals used for training. This type of training is supervised because labels are available for the training data. 1D-CNN are able to get trained to recognize different time domain signals with a small number of layers. Thus, the training can be performed efficiently even on standard cpu computers.

Once the 1D-CNN has been trained (i.e. all filter weights have been determined) then passing any candidate time domain signal through the 1D-CNN assigns a label to the signal that indicates what the signal is. 1D-CNN have been used for predicting defects in operating engines and motors [2, 4, 9], for diagnosing faults in bearing systems [6, 7, 8], and for identifying damage and the associated level in frame structures [10, 11]. In all of these cases labeled data must be available for all possible operating conditions at both pristine and damaged states. Additionally, any candidate signal considered in the future is placed in one of the available classes that are present in the training data. The supervised 1D-CNN need labeled data for all operating conditions and for all possible modes of fault which are expected to be identified. Such information requires a lot of effort and resources to generate which is not always practical. The new development presented in this paper aims in addressing this issue.

The "training" (i.e. computing the filter weights in the feature extractor + classifier) is based on how well they can recognize "labels" (i.e. operating speed, state of damage, etc.); a large amount of available labeled data (pristine + conditions of variable damage) is needed. Once trained the 1D-CNN can classify the level of damage associated with any candidate signal

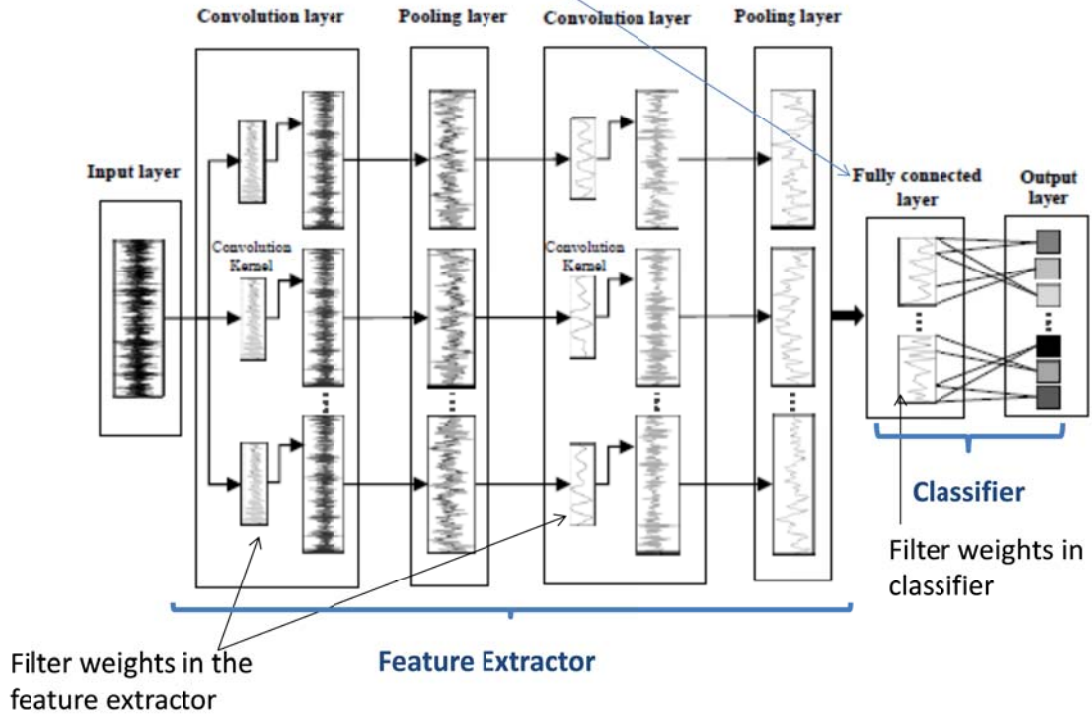

Figure 1. Operation of 1D-CNN (material from [18] is used for generating this Figure)

## 3. Temporal Clustering Network
The TCN presented in this paper first conducts the training in an unsupervised manner (i.e. no labeled data are required) and once trained, it is capable of recognizing when a signal from fault conditions is encountered (without determining the level of fault). Figure 2 presents the notional operation of the new capability. The clustering operation of the trained TCN uses the threshold and the failure probabilities computed for each cluster during the training for recognizing when a fault signal that does not belong to any of the clusters of pristine operating conditions is encountered.

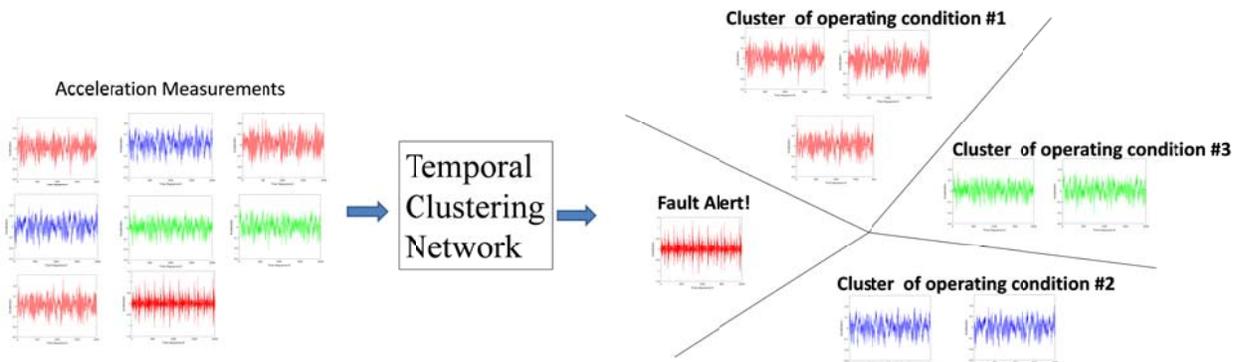

Figure 2. Representative operation of the developed TCN capability in identifying fault signals (the signals in the Figure are from the case study presented in this paper)

The TCN is comprised by the elements presented in Figure 3 and the unsupervised training takes place in the three steps presented in Figure 3. During the training it is expected that unlabeled time domain signals from all pristine operating conditions are available. The operations and the purpose of each Step are discussed next.

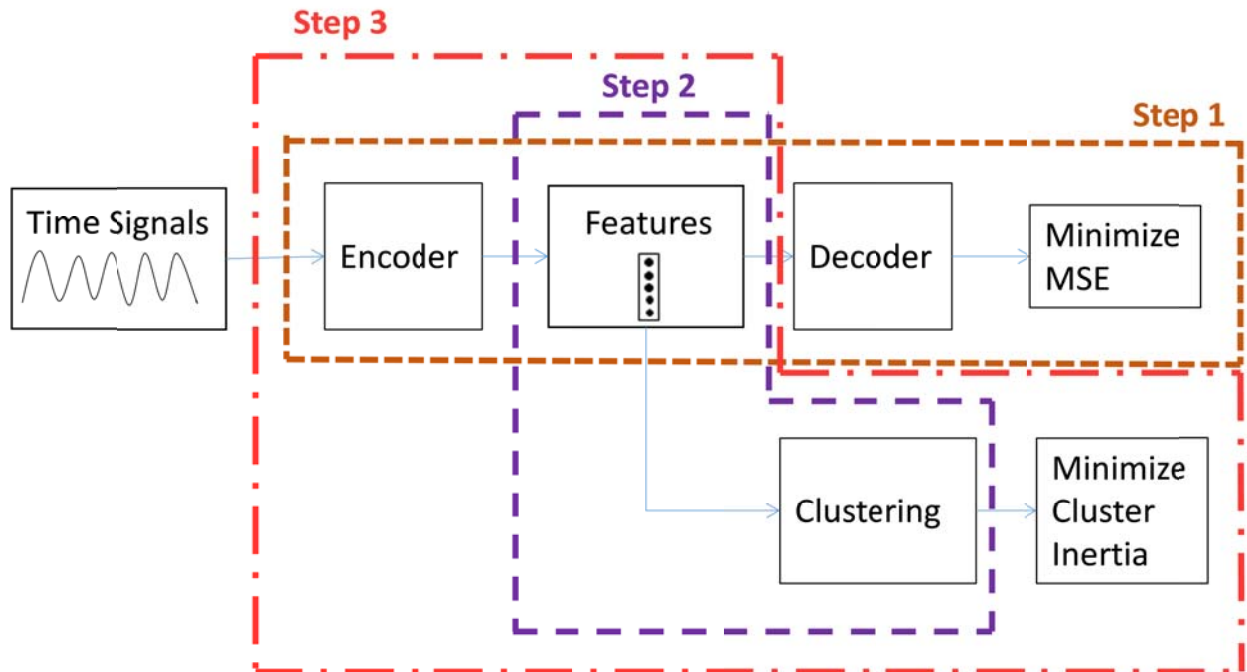

Figure 3. Steps in unsupervised training of the TCN process

**Step 1**. The purpose of this Step is to create values for the filter weights of the encoder so that any input time signal can be processed for generating its features. This capability is used by Step 2 for performing the initial clustering. The filter weights of the encoder and of the decoder are computed. Their values are determined based on how well the decoder recreates the original input time domain signal. During this step all available signals pass through the elements associated with Step 1 in Figure 3. The encoder operates in the same way with the feature extractor of 1D-CNN in Figure 1. It converts the time signal into a more compact representation while retaining most of the relevant information. This dimensionality reduction is necessary in order to avoid very long sequences which can lead to poor performance. The decoder recreates each input signal. By minimizing the mean square error between all input and all reconstructed signals it is ensured that the input time domain signals are well represented after the dimensionality reduction takes place in the encoder. The feature extraction capability of the trained encoder is utilized when conducting the initial clustering during Step 2. If the reconstruction quality is good, it is expected that the encoder has extracted the most relevant features from the input time signals. Such features include common features shared by all clusters and also features that are distinct to each cluster. The training is a straight forward optimization to minimize the cumulative reconstruction error. Therefore, the loss function for Step 1 training is defined as the mean square error between the reconstructed time signals and original time signals:

$$MSE = \frac{1}{I}\sum_{i=1}^{I}\|x_i - \hat{x}_i\|^2 \qquad (1)$$

where $I$ is the number of total signals used in the training (they are vibration measurements collected during the various operating conditions at the pristine state of the system); $x_i$ represents each original time signal, and $\hat{x}_i$ represents each reconstructed signal. The implementation in Python is using the capabilities provided by PyTorch; specifically the following functionalities were used: 1D convolutional layer, the 1D deconvolutional layer, the leaky-ReLU, the max pooling, and the max unpooling. The gradient calculations and the back propagation are handled automatically and efficiently by PyTorch. Figure 4 presents, as a sample, two sets of original and reconstructed signals at the conclusion of Step 1, corresponding to two of the vibration measurements which are used in the case study presented in Section 4. The results are representative of the level of accuracy achieved by the trained encoder/decoder. They are also representative of the similarities/differences of the vibration measurements collected during two different operating conditions.

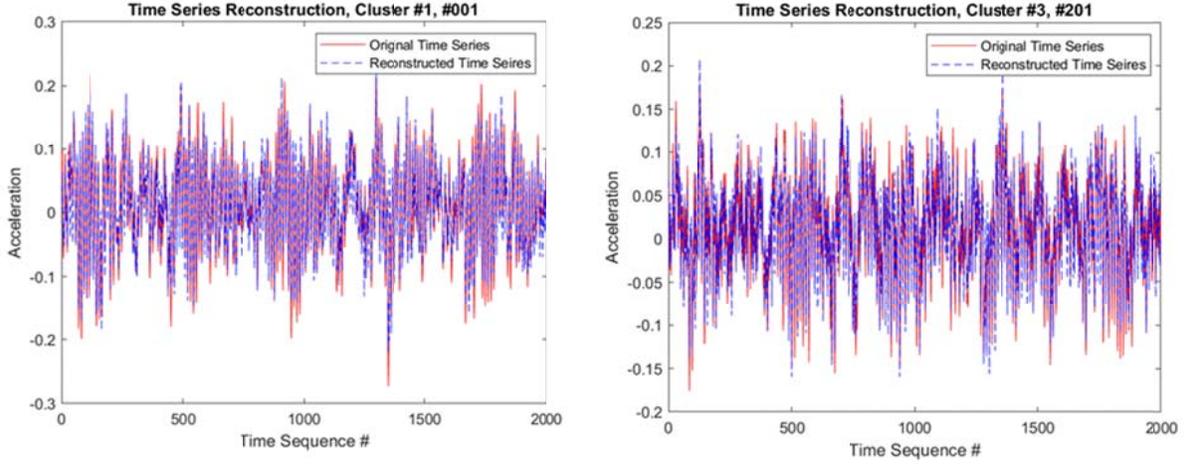

Figure 4. Samples of two sets of original and reconstructed signals at the conclusion of Step 1 of the TCN

**Step 2**. An initial clustering is completed by this Step. The filter weights of the encoder evaluated from Step 1 are used for generating the features of all available signals and the k-means clustering algorithm is used for processing the features and dividing the corresponding time domain signals into clusters. The purpose of the clustering in Step 2 is to place the time signals which are used for the training into clusters, and to also determine an initial set of values for the centroids of the clusters. The k-mean algorithm is iterative and partitions the dataset of time signals into sets of distinct non-overlapping clusters. The clustering in Step 2 tries to make the intra-cluster data points as tight as possible while also keeping the difference among clusters as large as possible. The following functional $J$ is minimized:

$$J = \sum_{i=1}^{I} \sum_{k=1}^{K} w_{ik} \|\boldsymbol{f}_i - \boldsymbol{\mu}_k\|^2 \tag{2}$$

where $K$ is the total number of clusters considered (in our work this value is set equal to the number of operating conditions which are encountered when collecting vibration measurements), $w_{ik}$ is an index which associates the $i^{th}$ signal with the $k^{th}$ cluster (it acquires the value of 1 when the signal $i$ gets assigned to the cluster $k$ and zero otherwise), $\boldsymbol{f}_i$ is the feature vector of the time signal $\boldsymbol{x}_i$, and $\boldsymbol{\mu}_k$ is the centroid of cluster $k$ in the feature space. The implementation of the initial clustering that takes place in Step 2 makes use of the k-means functionality which is available through the Scikit-Learn library. The outcome from Step 2 are the indices $w_{ik}$ for each training signal and the values $\boldsymbol{\mu}_k$ for the centroids.

**Step 3**. The final values of the filter weights of the encoder and of the centers of the clusters are computed. Information generated by the sequential execution of Steps 1 and 2 for the values of the filter weights of the encoder, the grouping of the training signals into clusters, and the values for the centroids of the clusters are used in Step 3. The computations in Step 3 re-calculate the filter weights of the encoder and the centers of the clusters by adjusting them in order to minimize the clustering inertia. In this step the clustering is no longer done by the k-means algorithm, but each time signal is clustered by the distance of its features to the cluster centroids. Additionally, the coordinates of each centroid are no longer calculated as the mean of each cluster, but their values are directly selected by the optimization process that minimizes the clustering inertia $CI$. Steps 1 and 2 are necessary before proceeding with Step 3 in order to expedite convergence and increase accuracy of the TCN process; they provide a good starting set of values for the filter weights of the encoder and for the coordinates of the centers of the clusters to Step 3 instead of starting with completely random values. Further, they provide the assignment of the training signals into the $K$ clusters; this assignment remains intact during Step 3. The clustering inertia is defined as:

$$CI = \sum_{k=1}^{K} \sum_{i=1}^{It_k} \|\boldsymbol{f}_i - \boldsymbol{\mu}_k\|^2 \tag{3}$$

where $It_k$ is the number of time signals in cluster $k$ (these values are determined from the indices $w_{ik}$). The implementation of the computations conducted in Step 3 is done using the functionality available by the PyTorch library; the clustering itself is explicitly coded in order to use the autograd capability for computing the gradients of $CI$ with respect to the filter weights of the encoder and the cluster centroids $\mu_k$. The final set of values for the filter weights of the encoder and the centroids of each cluster comprise the outcome of Step 3. Once this information is known, the training of the TCN is concluded.

**Utilization of trained TCN**

After the clustering has been completed at the end of Step 3, a threshold probability and a failure probability are defined for each cluster. Currently the threshold probability of each cluster is associated with the lowest 90% probability of the training signals placed in the cluster. It represents a low bound of probability for the training data to be recognized as a member of a particular cluster. Then, the failure probability is defined as a percentage (in this case 60%) of the threshold probability. It is used as a metric for comparing any candidate vibration signal. Any candidate vibration measurement collected during the operation of the system for which the TCN was trained, is processed through the encoder (using the filter weights from Step 3) and its features get computed. Then they are employed by the clustering (using the coordinates for the center of each cluster evaluated during Step 3) for evaluating the probability that the candidate signal is a member of each cluster. The latter is compared with the failure probability of the corresponding cluster. If the probability of a candidate signal processed by the trained TCN acquires a value lower than the failure probabilities of all clusters, it is identified as a signal containing a fault. It is envisioned that when a large percentage of fault signals are encountered within a group of candidate signals, then a fault alarm will be activated.

## 4. Case Study

In order to demonstrate how the TCN operates in recognizing fault signals, available vibration measurements at the bearings of an operating electric motor from [17] are used. Figure 5 presents the motor used for collecting the measurements (from [17]). Measurements for pristine conditions at 1,797 RPM; 1,770 RPM; and 1752 RPM are used for creating 120 time history signals for each operating condition (100 used for training and 20 used later in the validation). In Figure 2 just a few of those time signals are presented (in red for 1,797 RPM; blue for 1,770 RPM; and green for 1752 RPM). Without assigning a label (i.e. from which operating condition each time history originates) 300 training time signals comprised the unlabeled signals used for the unsupervised training of the TCN process (Figure 3). After the training has been completed, as expected, the TCN sorts the 300 training time signals into three clusters as depicted in Figure 6. For information purposes, although labels were not used for the training, by inspecting the way that the clusters were formed, only one signal from the 1,770 RPM conditions was clustered together with the signals from the 1,797 RPM conditions. The rest of the signals ended up in the expected cluster.

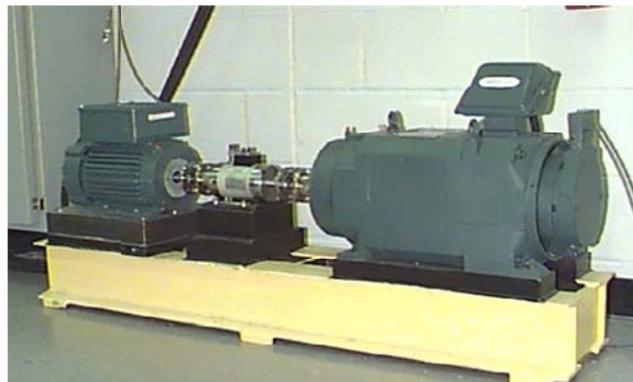

Figure 5. Motor and measurement stand used in [17] for collecting vibration measurement under pristine and fault bearing conditions

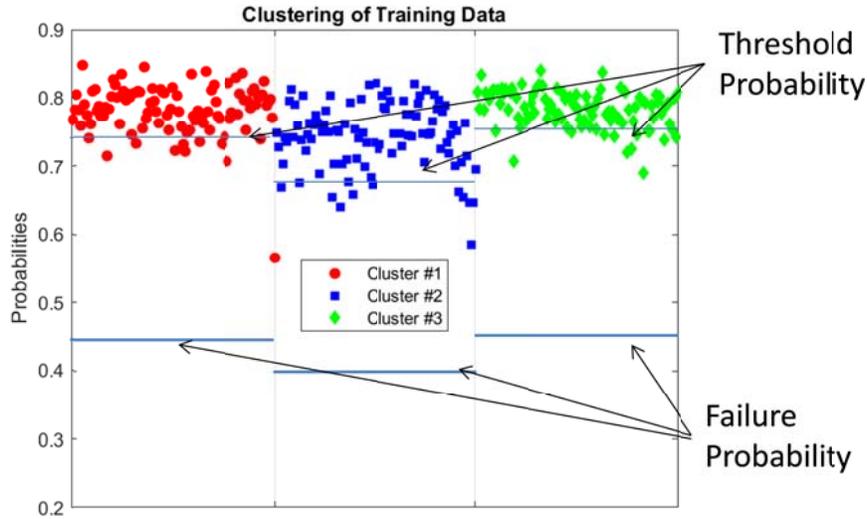

Figure 6. Clustering of training data and definition of threshold and failure probabilities (red used for signals placed in the cluster associated with 1,797 RPM; blue for signals in the cluster associated with 1,770 RPM; green for signals in the cluster associated with 1,752 RPM)

The results are used for determining the threshold and the failure probability values for each cluster (presented with solid horizontal lines in each cluster). Once the TCN has been trained and the failure probabilities are available, 20 time signals at pristine conditions from each one of the three RPM and 20 time signals from the lowest state of bearing damage available in [17] (a dent at the bearing inner race, with diameter 0.007" and depth 0.011") at each one of the three operating RPM (a total of 120 time histories used as validation data) are processed by the TCN. Results for the clustering of the 60 validation data from pristine conditions are presented in Figure 7; the probabilities assigned to all 60 signals at each cluster are presented. None of these 60 signals were used during the unsupervised training. As long as the probability of a signal from pristine condition is above the failure probability in one of the clusters, then it is correctly identified as a signal from bearings with no fault. Out of the 60 signals 10 of them (it happens that all are from the 1,770 RPM (blue)) are erroneously identified as signals from faulty bearings.

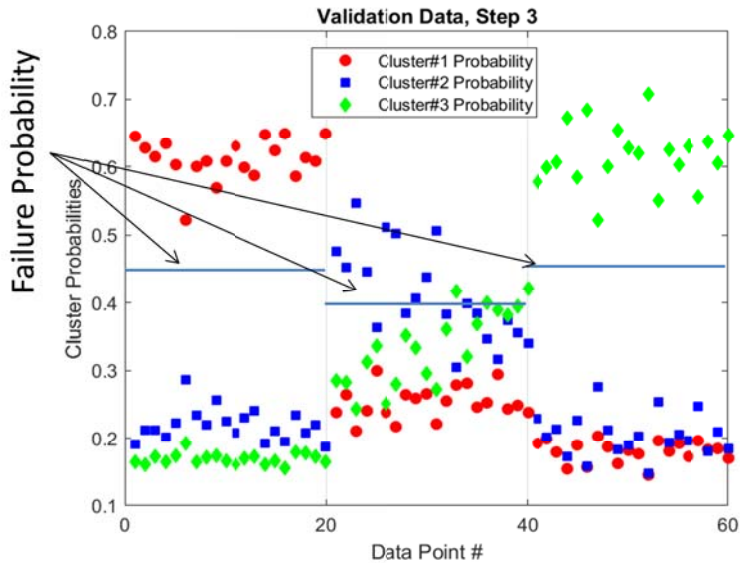

Figure 7. Clustering of validation signals from pristine state of bearings

Results for the clustering of the 60 validation data from the fault conditions are presented in Figure 8. The probabilities assigned to all 60 signals are presented in each cluster, and all of them exhibit a probability of belonging to any one of the three clusters which is far below the associated failure probabilities. Therefore, all of them are correctly recognized as signals from a condition with a fault. Overall, out of the 120 validation signals only 10 signals are not identified correctly. We did not use the available margin to adjust the failure probability in order to identify correctly more pristine signals during the validation.

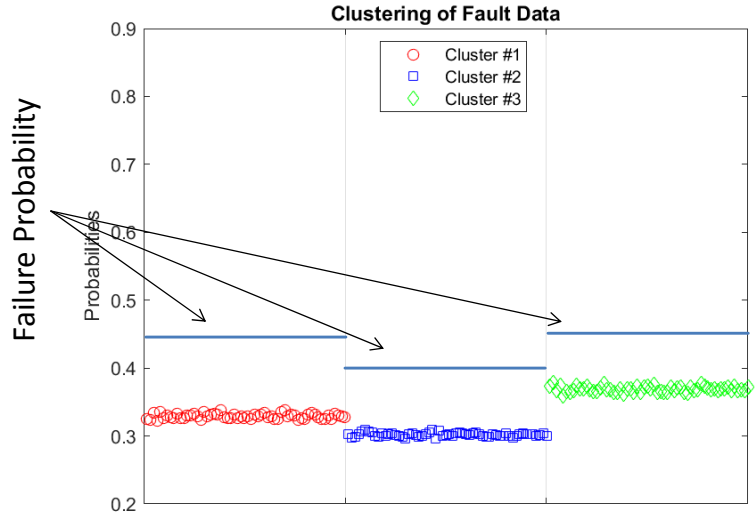

Figure 8. Clustering of fault data and placement well below the failure probability values

## 5. Suggested Future Research
The work presented in this paper introduces the TCN as a meaningful approach for self-diagnosing faults in an operating system through unsupervised training. The following work can further solidify and validate the new approach.
a)  Create a large set of experimental data under both steady-state and transient loading conditions at pristine and damaged states of the operating structure. A frame structure made out of beams that are connected with bolt joints (similar to [10, 11]) and excited by a shaker is a good candidate structure because various levels of damage can be introduced by loosening one bolt at a time. It is also easy to construct and test.
b)  Investigate the following aspects of the unsupervised training: (i) The number of layers in the encoder. (ii) The number of filters at each layer. (iii) The kernel size (it determines the number of weights comprising each filter, and the dimensional reduction achieved in each layer during the associated maxpool operation). (iii) The stride size (determines how densely the input time signal in each layer is sampled). (iv) The definition of the loss function that drives the unsupervised training. (v) An iterative algorithm that decides concurrently the filter weights of the encoder, the placement of the signals into clusters, and the centers of the clusters, while avoiding trivial solutions during the creation of the clusters. This work will establish guidelines on how to construct, define, and operate the TCN for processing vibration signals under both steady state and transient loading, and being able to recognize a variable stage of damage.
c)  Solidify through statistical methods the determination of the threshold probability and of the failure probability for each cluster. Statistical methods that identify outliners in the training data will be employed for determining the threshold probability in a mathematically rigorous manner (instead of selecting the 90% lowest probability value in the presented work). Further, statistical methods for set-membership estimation will be used for determining the failure probability for each cluster (instead of selecting the 60% of the threshold probability value as done in the presented work).
d)  Establish guidelines about the time duration of both the training signals and the candidate signals processed by the trained TCN. The guidelines should consider the dynamic characteristics of the system and the operating conditions of interest. Further, criteria should be established on the percentage of fault activations encountered within a pre-determined number of sequential signals before a fault alarm is generated. Such a capability is necessary for avoiding false fault alarms when statistical outliers are encountered.

e) Develop an on-site and repeatable unsupervised training process for updating the statistics and the definition of the clusters (this capability will be needed for maintaining the trained TCN up-to-date when the mission profile changes, equipment gets upgraded, etc.).